\documentclass[prl,preprint,showpacs,preprintnumbers,amsmath,amssymb,epsfig]{revtex4}

\usepackage{bm}% bold math
\usepackage{graphicx}% Include figure files

\begin{document}

\title{Superradiant scattering from a hydrodynamic vortex}
\author{T.R. Slatyer and C.M. Savage}
\email{craig.savage@anu.edu.au}
\affiliation{Australian Centre for Quantum Atom Optics, Australian National University, ACT 0200, Australia}

 \begin{abstract}
We show that sound waves scattered from a hydrodynamic vortex may be amplified.  Such superradiant scattering follows from the physical analogy between spinning black holes and hydrodynamic vortices.  However a sonic horizon analogous to the black hole event horizon does not exist unless the vortex possesses a central drain, which is challenging to produce experimentally.  In the astrophysical domain, superradiance can occur even in the absence of an event horizon: we show that in the hydrodynamic analogue, a drain is not required and a vortex scatters sound superradiantly. Possible experimental realization in dilute gas Bose-Einstein condensates is discussed.
\end{abstract}


\date{\today}
\maketitle

Hawking's prediction of radiation from black holes is one of the most surprising outcomes of the quest to unify quantum mechanics and general relativity \cite{hawking}. It is problematic that such an important theoretical prediction has no immediate prospect of direct experimental investigation. Consequently, Unruh's observation of an analogy between wave scattering from black holes, and sound wave scattering from sonic horizons in fluids, has intrigued both the astrophysical and condensed matter communities \cite{unruh,zurek,visser,jacobson}. 

The generation and detection of the sonic analogue of Hawking radiation faces many difficulties, and a detailed experimental scheme has yet to be presented. Consequently, we propose the initial investigation of the simpler, but closely related, superradiant amplification of scattered waves by an ergoregion \cite{bekenstein}.   Quantum mechanically, Hawking radiation and superradiance may be regarded as spontaneous and stimulated radiation counterparts. However, stimulated radiation is easier to observe because of its much greater strength, and because of the control provided by the stimulating pulse. Furthermore, an analogue ergoregion is realized by a vortex, which is experimentally simpler to achieve than a horizon. This superradiance is related to the Zeldovich-Starobinsky effect, which has been discussed in the context of superfluid helium by Volovik \cite{volovik}. We provide quantitative details and numerical simulations to argue that this superradiance might be observed in dilute gas Bose-Einstein condensates (BECs) \cite{garay,leonhardt,barcelo,giovanazzi}.

According to general relativity, rotating black holes have an ergoregion: a region in which no physical object can remain at rest with respect to an inertial observer at infinity.  Low frequency waves are scattered from the ergoregion with increased amplitude; that is superradiantly \cite{frolov}.  In fluids, regions where the fluid speed exceeds the speed of sound are analogue ergoregions  \cite{visser}. A draining vortex with a central sink may also possess an analogue event horizon, and hence is an analogue of a rotating black hole.  Basak and Majumdar \cite{basak} showed that sonic superradiance occurs for a draining vortex, but assumed that the density of the fluid, and the speed of sound, were constant.

For a BEC a sink might be realized by outcoupling the condensate to form an atom laser beam \cite{bloch}. However there are difficulties with strong outcoupling \cite{robins}, and it will be simpler to create a vortex without a sink, for example by  stirring with laser beams \cite{madison} or magnetic fields \cite{hodby}.  Then the simpler velocity function makes it feasible to consider the variable density profiles which are unavoidably associated with vortices in BECs.

The wave equation for sound waves, that is linear perturbations of the velocity potential $\phi$ of a barotropic, inviscid, irrotational fluid, may be written \cite{unruh,visser}
\begin{equation} 
\partial_{\mu} \left( g^{\mu \nu} \sqrt{-g} \partial_{\nu} \phi \right) = 0 \; ,
\label{kleingordon} 
\end{equation}
where the coefficients $g^{\mu \nu}$ are functions of the unperturbed fluid density and velocity, and $g$ is their inverse determinant. This has the same form as the Klein-Gordon equation for a massless scalar field in curved spacetime with metric $g_{\mu \nu}$ \cite{birrell}. Provided the quantized fluid has suitable commutation relations, the analogy extends to the quantum field \cite{unruh and schutzhold}.

All this suggests the possibility of using fluid flows to mimic general relativistic spacetime metrics.  The analogy requires the length scales of the problem to be large enough that the atomicity of the fluid can be neglected.  In BECs, furthermore, the hydrodynamic equation (\ref{kleingordon}) is an approximation neglecting the ``quantum pressure'' term \cite{pethick}.  This is valid when any spatial variations in the BEC occur over length scales longer than the healing length of the condensate  \cite{liberati}. In our case, this must be smaller than both the sound wavelength and the spatial scale of the vortex. 

Let the unperturbed fluid's speed of sound be $c$, velocity  be $\vec{v}$, and density be $\rho$.  In cylindrical polar coordinates a vortex centred on the origin, with $v_r = v_z = 0$, has wave equation coefficients
\begin{equation} 
g^{\mu \nu} = \frac{-1}{\rho c} 
\left( \begin{array}{cccc}
 1 & 0 &  v_{\theta}/r &0 \\ 
 0 &  -c^2  &  0 & 0 \\ 
 v_{\theta}/r & 0 & (v_{\theta}^2 -c^2 )/{r^2} & 0 \\ 
 0 &  0 &  0 & -c^2
  \end{array}  \right) \; ,
\end{equation}
and $g = - \rho^4  r^2 / c^2$. For irrotational flow, $v_{\theta} = \alpha/r$, for some constant $\alpha$.  The density and speed of sound depend only on the radial coordinate $r$, and approach asymptotic values $\rho_{\infty}$ and $c_{\infty}$ for $r \rightarrow \infty$. 

We consider cylindrical wave solutions to the wave equation (\ref{kleingordon}) of the form $\phi (t, r, \theta, z) = \psi(t,r) e^{-i m \theta}$, with angular wavenumber $m$.  Assuming that the square of the speed of sound is proportional to the density, as is the case for a BEC \cite{pethick}, the density may be eliminated from the wave equation, which then becomes
\begin{equation} 
\frac{\partial^2 \psi}{\partial t^2} -2i \frac{m v_\theta}{r} \frac{\partial \psi}{\partial t} -
\frac{1}{r} \frac{\partial}{\partial r} \left( r c^{2} \frac{\partial \psi}{\partial r} \right) 
+ \frac{m^{2}}{r^2} \left( c^{2} - v_\theta^{2} \right)  \psi  
= 0 \; .
\label{wave equation}
\end{equation}
For single frequency waves of the form $\psi (t, r) = r^{-1/2} G(r^*)  e^{i  \omega t }$, and defining a ``tortoise coordinate'' $r^*$ by $ dr/d r^* = \tilde{c}^2$, with $r^* \rightarrow r$ for $r \rightarrow \infty$, where $ \tilde{c} = c / c_{\infty} $, we find
\begin{align} 
& \frac{d^2 G(r^*)}{dr^{*2}} + \frac{\tilde{c}^2}{c_{\infty}^2} 
( \omega^2 - V_{\text{eff}} ) G(r^*) = 0 \; ,\label{G de} \\
& V_{\text{eff}} =   \frac{2 m \omega v_\theta}{r} + \frac{m^2}{r^2} \left( c^2 - v_\theta^2 \right) - \frac{1}{2r} \left( \frac{c^2}{2r} - \frac{dc^2}{dr} \right)  \; ,
\label{effective potential}
\end{align}
where we have introduced an effective potential $V_{\text{eff}}$ to emphasise the similarity to the time-independent Schr\"{o}dinger equation.

In order to analyse superradiance we consider the two limiting cases where the tortoise coordinate $r^*$ approaches $\pm \infty$.  For $r$ large, $r^* \approx r$, and eliminating all terms except those of highest order in $r$ yields the asymptotic form
\begin{equation} 
\frac{d^2 G(r^*)}{dr^{*2}} + \frac{\omega^2}{c_{\infty}^2} G(r^*) \approx 0 \; ,
\end{equation}
with general solution
\begin{equation} 
G(r^*) = A e^{i(\omega / c_{\infty} ) r^*} + B e^{-i(\omega / c_{\infty} ) r^*} \; ,
 \label{posinfty} 
\end{equation}
where $A$ and $B$ are constant amplitudes of incoming and outgoing waves respectively. 

The analysis of the other limit, where $r^* \rightarrow - \infty$, requires that the density profile of the vortex be specified.  In order to present analytical calculations we use the density profile
\begin{equation}
 \rho (r) = \rho_{\infty} \frac{\left[(r - r_{0})/\sigma\right]^2}{2 + \left[(r - r_{0})/\sigma \right]^2} \; .
 \label{density profile}
 \end{equation}
This is similar to the charge $l=1$ vortex density profile for a BEC  \cite{pethick}, but with the scale length given by the free parameter $\sigma$, rather than by the healing length $\chi$.  In the Thomas-Fermi limit of dominant atom interactions, the length scale for a charge $l$ vortex, with angular momentum $l \hbar$ per particle, is $l \chi$ \cite{svidzinsky}. Although a vortex with $l > 1$ is unstable to decay into $l$ single charge vortices, it may be stabilised by a pinning optical potential \cite{savage}. Since the density profile Eq.(\ref{density profile}) was chosen primarily for analytical convenience, we do not expect it to be exactly achievable experimentally, although something close to it should be. For example, the dipole potential generated by far detuned light has been used to engineer density profiles in BECs \cite{stamper-kurn}. The other free parameter is the radius of the zero density core $r_0$, which might also be engineered by applied potentials. Since we have assumed that the square of the speed of sound is proportional to the density, $\tilde{c}^2 = \rho / \rho_{\infty} $, and the tortoise coordinate is
\begin{equation} 
r^* = \int \frac{1}{\tilde{c}^2} \, d r = r - \frac{2 \sigma^2}{r - r_{0}} \; .
\label{tortoise}
\end{equation}
For $ r \approx r_{0} $, $r^* \approx - 2 \sigma^2 / (r - r_{0}) $ and $ c^2 \approx c_{\infty}^2 (r - r_{0})^2 / (2 \sigma^2) $.  Using these approximations and retaining only terms of lowest order in $ r - r_{0} $, we find that asymptotically
\begin{align} 
& r^{*2} \frac{d^2 G(r^*)}{dr^{*2}} + \frac{2 \sigma^2 \Omega^2}{c_{\infty}^2} G(r^*) = 0 \; ,  \\
& \Omega \equiv \omega - {m \alpha}/{r_{0} ^2} \label{Omega}  \; .
\end{align}
This is an Euler-type equation and has solutions of the form $G(r^*) = |r^*|^{\delta} $, where
\begin{equation} 
\delta = \frac{1}{2} \pm \gamma ,\quad
\gamma \equiv \frac{1}{2} \mathrm{sign}(\Omega) (1 - 8 \sigma^2 \Omega^2 / c_{\infty}^2)^{1/2} \; .
\label{delta} 
\end{equation}
The general solution to the asymptotic form of the wave equation is then
\begin{equation} 
G(r^*) = | r^* |^{1/2} \left( C | r^* |^{\gamma} + D | r^* |^{- \gamma} \right)  \; ,
\end{equation}
where $C$ and $D$ are constants. According to Eq.(\ref{delta}), $\gamma$ is either real or purely imaginary. For $\gamma$ imaginary this asymptotic solution is oscillatory and may be written in the form
\begin{equation} 
G(r^*)  = | r^* |^{1/2}  \left( C e^{ \gamma  \ln{ | r^* |} } 
+ D e^{- \gamma  \ln{ | r^* |} } \right) \; .
\label{neginfty} 
\end{equation}
By making $| r^* |$ the argument of a logarithm we have anticipated that $r^*$ will be made dimensionless.  This last form allows us to identify the two linearly independent solutions as outgoing and ingoing waves, with amplitudes $C$ and $D$, respectively.  Note that in this asymptotic limit $r^* < 0$ and hence $\ln{ | r^* |}$ increases with decreasing $r^*$. Note also, that it is the sign of the group velocity, not of the phase velocity, which determines the identification of outgoing and ingoing waves \cite{vilenkin}, and the sign of the group velocity is independent of the sign of $\Omega$.

Let the Wronskians (of the solution and its complex conjugate) corresponding to the asymptotic solutions Eqs.~(\ref{posinfty}, \ref{neginfty}) be denoted $W(\infty)$ and $W(- \infty)$. Then
\begin{equation} 
W(\infty)  =  \frac{2 i \omega}{c_{\infty}} \left( |B|^2 - |A|^2 \right) \; .
\label{Wronskian+inf}
\end{equation}
If $\gamma$ is imaginary, 
\begin{equation} W(- \infty) = 2 \gamma ( |C|^2 -|D|^2 )  \; . 
\label{Wronskian-inf-im}
 \end{equation}
Following Vilenkin \cite{vilenkin}, consider a solution to the wave equation for $G(r^*)$ representing a wave originating at $ r^* = + \infty$ and having the asymptotic forms, for imaginary $\gamma$,
\begin{equation} 
G(r^*) = \left\{ \begin{array}{ll} 
e^{i(\omega / c_{\infty} ) r^*} + R e^{-i(\omega / c_{\infty} ) r^*} &  \; , r^* \rightarrow \infty \\ 
T | r^* |^{1/2} e^{-\gamma \ln | r^* |} &  \; , r^* \rightarrow - \infty 
\end{array} \right.  \; .
\end{equation}
Then by Abel's Theorem, the Wronskian of this solution is constant, since there is no first derivative term in the differential equation for $G(r^*)$, Eq.~(\ref{G de}).  Thus, setting $A = 1$ and $B = R$ in Eq.~(\ref{Wronskian+inf}), and $C = 0$ and $D = T$ in Eq.~(\ref{Wronskian-inf-im}), and equating $W(\infty) = W(-\infty)$,
\begin{equation} 
\frac{2 i \omega}{c_{\infty}} ( |R|^2 -1 )  =  -2 \gamma |T|^2   \;  ,
\end{equation}
and therefore,
\begin{equation} 
|R|^2 = 1 -\mathrm{sign}(\Omega)  |\gamma| |T|^2 c_{\infty}/ \omega \; .
\end{equation}
In particular, if $\Omega$ is negative, and $T \neq 0$, then $|R| > 1$, and superradiance occurs, since then, in the asymptotic region far from the vortex, the amplitude of the reflected wave exceeds that of the incident wave. This superradiance inequality $\Omega < 0$ expands to $\omega < m \alpha / r_0^2$ , which requires $m \alpha > 0$, so that the cylindrical waves propagate in the same direction as the vortex flow.

In contrast to the draining vortex examined by Basak and Majumdar \cite{basak} and the rotating black hole \cite{vilenkin}, here there are two independent necessary conditions for superradiance: $\gamma$ must be purely  imaginary and $\Omega$ must be negative.  Combining these conditions gives a frequency-independent condition for superradiance. For since $\omega \geq 0$, then $\Omega \geq - m \alpha / r_0^2$, and superradiance requires, $ 0 \geq \Omega \geq - m \alpha / r_0^2$, and therefore $\Omega^2 \leq m^2 \alpha^2 / r_0^4 $.  If $\gamma$ is to be imaginary, we must have $1 < 8 \sigma^2 \Omega^2 / c_{\infty}^2$, and since $m \alpha > 0$, the frequency-independent necessary condition for superradiance is $\sigma m \alpha > c_{\infty} r_0^2 / (2 \sqrt{2})$.

For a charge $l$ vortex in a BEC we can substitute expressions \cite{pethick} for the vortex velocity constant $\alpha = l \hbar / m_{\text {atom}}$, where $m_{\text {atom}}$ is the mass of an atom, and for  $c_\infty$ in terms of the healing length $\chi$, $c_\infty = \hbar / ( \sqrt{2} \chi m_{\text {atom}} )$, to obtain $\chi > r_0^2 / (4 \sigma l m )$. This is always possible to fulfill, at least in principle, since the healing length is given by $\chi =(8 \pi n a)^{-1/2}$, where $n$ is the atomic number density and $a$ is the s-wave scattering length, and both $n$ and $a$ are under experimental control.
\begin{figure}
\includegraphics[width=\columnwidth]{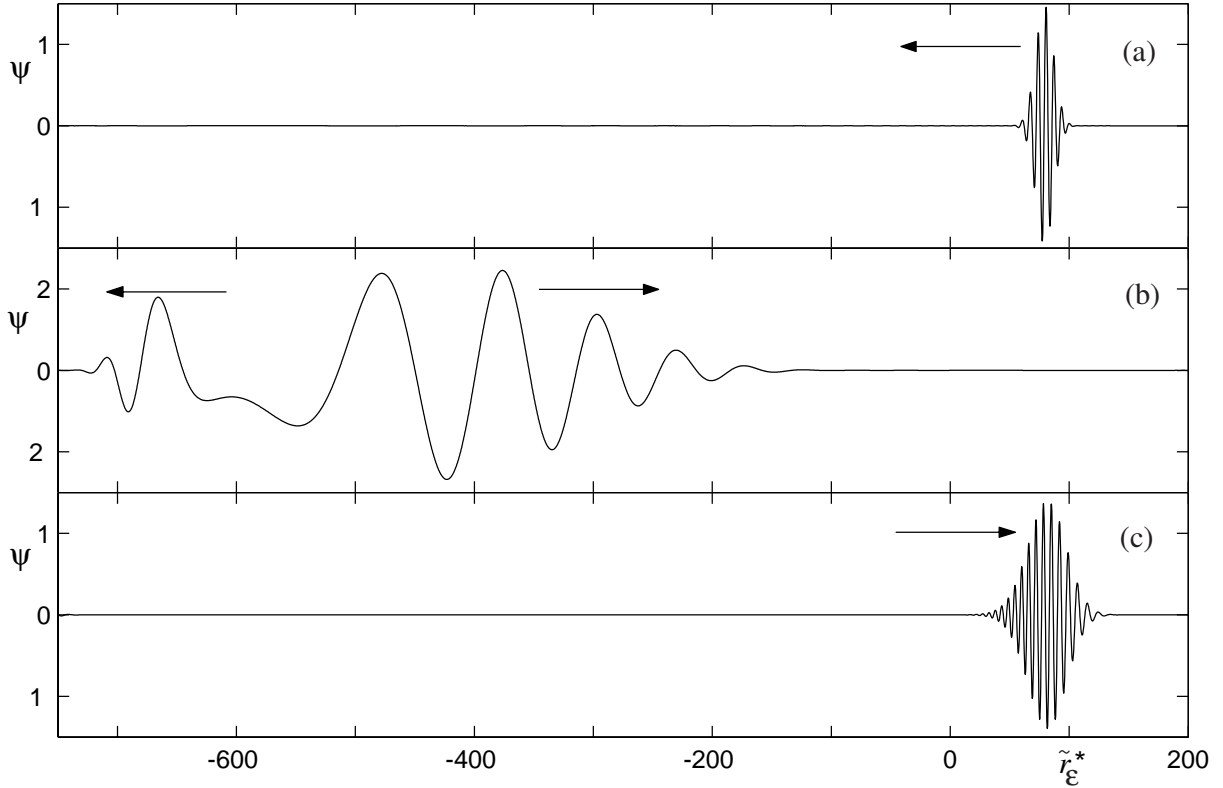}
\caption{Wavepacket propagation from numerical solution of Eq.~(\ref{wave equation}). The real part of the wave packet is plotted versus the dimensionless modified radial tortoise coordinate $\tilde{r}^*_\epsilon$, defined in the text, with $\epsilon =800/801$ so that $\tilde{r}^*_\epsilon = -800$ at $r=r_0$. The initial wavepacket, at $\tilde{t} = 0$, was given by Eq.(\ref{initial wavepacket}) with $\tilde{r}_{\text{init}} = 200$ and $\tilde{w} = 10$. (a) $\tilde{t} = 114$: the wavepacket is propagating towards the vortex. (b) $\tilde{t} = 330$: the wavepacket has just split into reflected and transmitted parts. (c) $\tilde{t} = 504$: the reflected wavepacket. Dimensionless parameters: $m = 1$, $\tilde{\alpha} = 2$,  $\tilde{c}_\infty = 1$, $\tilde{\sigma} = 20$. The spatial grid had 8192 points over $\tilde{r}^*_\epsilon \in [-800,400]$, and the time steps were 0.015 dimensionless time units. }
\label{wavepacket propagation}
\end{figure}

To determine the magnitude of the superradiant amplification, we numerically solved the wave equation (\ref{wave equation}) using the XMDS package \cite{xmds}. We used an initial Gaussian wavepacket of the form $\psi(0,r) = A(r) $, with frequency $\omega_0$. Specifically,
\begin{align} 
& A(r) = e^{-(r - r_{\text{init}})^2/w^2}  e^{ i \omega_0 r / c_\infty}
\; ,\nonumber \\
& \frac{\partial \psi}{\partial t} (0,r) =  \left( i \omega_0 - 2 c_{\infty} (r - r_{\text{init}}) / w^2 \right) A(r) \; .
\label{initial wavepacket}
\end{align}
We found that superradiance occured for a wide range of parameters, as long as the relevant inequalities were fulfilled. Fig.~\ref{wavepacket propagation} shows a particularly strong and clean example of the scattering of an $m=1$ wavepacket. Comparing the Fourier components of the real parts of the incident and reflected wavepackets in the asymptotic region, we found that the dominant Fourier power in the reflected wavepacket is approximately doubled. 

The parameters are given in the caption, and are representative of conditions for a trapped dilute gas BEC.  They are made dimensionless, indicated by a tilde, by measuring distance and time in units of $r_0$ and $\omega_0^{-1}$ respectively. For example, choosing $r_0 = 1$ $ \mu$m and $\omega_0^{-1} = 10^{-3}$ s, the parameters of Fig.~1 correspond to $c_\infty = 10^{-3}$ ms$^{-1}$ and $\sigma = 20$ $\mu$m. For the case of a Rb$^{87}$ BEC, $m_{\text {atom}} = 1.4 \times 10^{-25}$ kg, and the asymptotic healing length $\tilde{\chi} \approx 0.5$, and $\tilde{\alpha} =2$ corresponds to a vortex charge of about $l=3$.    Note that the real part of the wave is plotted against the modified tortoise coordinate $\tilde{r}^*_\epsilon = \tilde{r} - 2\tilde{\sigma}^2 / ( \tilde{r} -1 + \epsilon )$. This magnifies the scale as $r \rightarrow r_0$ $(\tilde{r} \rightarrow 1)$ , while limiting the lower bound of $\tilde{r}^*_\epsilon$ to $1 -2\tilde{\sigma}^2 / \epsilon$. The ergoregion boundary, or static limit, is at $\tilde{r} \approx 8$ or $\tilde{r}^*_\epsilon \approx -92$.

The local healing length determines the validity of the hydrodynamic approximation, and is larger than the asymptotic value by the factor $\sqrt{ \rho_\infty / \rho (r) }$. Because of this the BEC interpretation fails for Fig.~1(b). However, we have found a tradeoff between the strength of the superradiance and the validity of the BEC hydrodynamic approximation. With increasing $m$, $\tilde{\alpha}$, and sound wavelength, reflection occurs at larger values of $r$, and hence for smaller local healing lengths, while the superradiance decreases in strength. We define the reflection point to be at the maximum of the effective potential Eq.~(\ref{effective potential}). Compared to the case of Fig.~1, a transmitted wavelength comparable to the local healing length at the reflection point, accompanied by a 70\% reduction in superradiance power, occurs for an initial wavelength an order of magnitude longer, and $\tilde{\alpha}$ a factor of five larger. 

Numerical simulations of BECs under the proposed experimental conditions, and without the hydrodynamic approximation, could explore the effect of its breakdown as the wavelength of the transmitted wave decreases due to its inward propagation. This is related to the ``trans-Planckian'' problem of Hawking radiation \cite{unruhprd}.

In summary, our analytical and numerical work has shown superradiant sound scattering from hydrodynamic vortices. We have also argued for its experimental realization in dilute gas Bose-Einstein condensates. This might provide a useful step towards the ultimate goal of observing the sonic analogue of Hawking radiation.

ACQAO is an Australian Research Council Centre of Excellence. This research was supported by the Australian Partnership for Advanced Computing.

\end{document}